# Molecular dynamics simulation of fabrication of Cu mono-component metallic glass by physical vapor deposition on Zr substrate

Yang Yu*, Yunyun Chen, Fenping Cui

*Nanjing University of Information Science and Technology, School of Physics and Optoelectronic Engineering, China*

*Authors to whom correspondence should be addressed: Dr. Yang Yu, Electronic mail: yuyang5020@googlemail.com

Abstract:

In this work, the single-component Cu metallic glass was fabricated by the physical vapor deposition on the Zr (0001) crystal substrate at 100 K using the classical molecular dynamic simulation. The same deposition process was performed on the Cu (1 0 0) and Ni (1 0 0) crystal substrate for comparison, only the Cu crystal deposited layer with the fcc structure can be obtained. When depositing the Cu atoms on the Zr substrate at 300 K, the crystal structure was formed, which indicates that except the suitable substrate, low temperature is also a key factor for the amorphous structure formation. The Cu liquid quenching from 2000 K to 100 K were also simulated with the cooling rate $10^{12}$ K/s to form the Cu glass film in this work. The Cu metallic glass from the two different processes (physical vapor deposition and rapid thermal quenching from liquid) revealed the same radial distribution function and X-ray diffraction pattern, but the different microstructure from the coordination number and Voronoi tessellation analysis.

**Introduction:**

Since the first metallic glass was successfully fabricated in 1960[1], many techniques have been developed to impel the manufacture of metallic glass. Plentiful new multicomponent metallic glasses with different components have emerged. The metallic glasses exhibit many attractive properties such as high strength, high hardness and high corrosion resistance. Meanwhile, they possess good ductility and toughness [2-6]. Compared to the developed technique of multicomponent metallic, due to the low-glass forming ability of pure glass metal, which needs ultrahigh cooling rate, the fabrication of single-component metallic glass was only realized in computer simulation[7-10]until recently. The experimental realization of single element metallic glass was accomplished in the reported works [11, 12]. In this work, the physical vapour deposition was supplied to synthesize pure metallic glass using molecular dynamics simulation to show an experimental feasible way for the manufacture of the pure metallic glass.

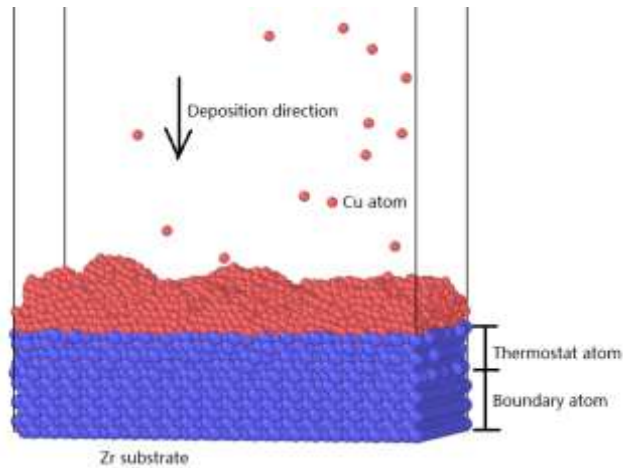

Figure 1 Snapshot of Cu atoms depositing on Zr (0001) crystal substrate with the downward velocity corresponding to the incident energy 0.09 eV.

**Simulation method:**

All the simulations performed in this work applied the LAMMPS (large-scale atomic/molecular massively parallel simulator) molecular dynamics package [13, 14] within the microcanonical NVE ensemble. The time step in the molecular simulation was 1 fs. The X and Y directions were periodic, and the Z axis was fixed. There were three kinds of substrate: Zr(0 0 0 1), Cu (1 0 0) and Ni(1 0 0). The dimension for the three substrate crystal were as follows: for Zr 64.64 Å ×111.96 Å × 26.39 Å; for Cu 72.30 Å ×72.30 Å ×18.07 Å; for Ni 70.48 Å ×70.48 Å ×17.62 Å. All the substrates comprised of the number of 8800 atoms. The bottom six layers of the substrates were boundary atoms, used to prevent atom loss; then the five layers up were thermostat layers. The substrate atoms were fixed to 100 K by langevin thermostat [15-18]. The Cu atoms were deposited on the three substrates with the downward velocity corresponding to the incident energy 0.09 eV. The deposition rate was one atom per ps. The deposited Cu thin film was formed with the total number of 8000 atoms. The deposited Cu layer can be separated from the substrates for further analysis and denoted as CuDepZr, CuDepCu, CuDepNi in this work. For comparison, the Cu liquid cooling from 2000 K to 100 K with the high cooling rate $10^{12}$ K/s was simulated in this work. The rapid quenching Cu film sample was named CuFilm in this work. The CuFilm comprised 16800 atoms of the dimension 72.30 Å ×72.30 Å ×36.15 Å.

The embedded-atom-method (EAM) potential [19, 20] was used in the LAMMPS MD package for the Cu-Zr system. Since the two-component potential file is also valid for either element for pure metal simulation, the same Cu-Zr EAM potential was also applied for the Cu-Cu system to make the comparison consistent. The Cu-Ni EAM potential[21] was applied for the Cu-Ni system.

The Ovito software [22, 23] was employed for the radial distribution function (RDF), coordination number and Voronoi tessellation [24] analysis. The X-ray diffraction (XRD) data was obtained from the LAMMPS USER-DIFFRACTION package [25] [26] [27].

**Result and discussion:**

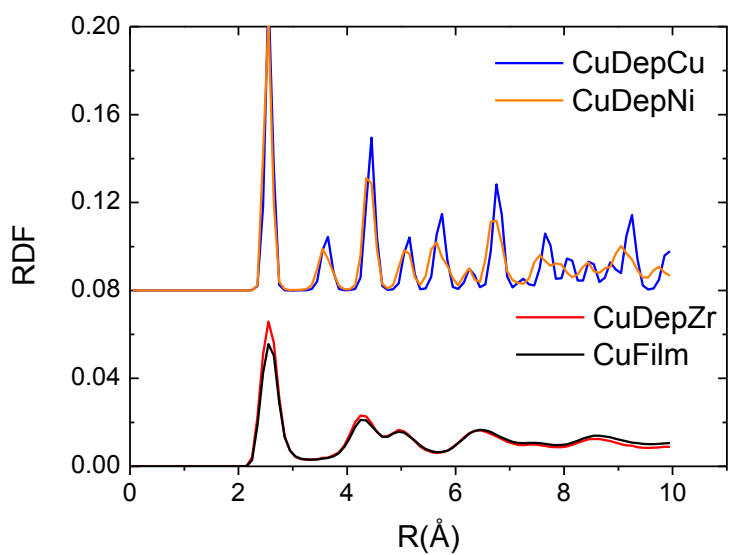

Figure 2 Radial distribution function (RDF) of Cu atoms depositing on Zr (CuDepZr), Cu (CuDepCu) and Ni (CuDepNi) crystalline substrates as well as the RDF of the rapid cooling of Cu liquid (CuFilm).

The radial distribution function (RDF) of Cu atoms depositing on Cu (CuDepCu), Ni (CuDepNi) and Zr (CuDepZr) crystalline substrate as well as the rapid cooling of Cu liquid (CuFilm) is shown in figure 2. The upper lines (blue and orange) which corresponds to the CuDepCu and CuDepNi reveal the long range order characteristic of crystals, confirm the crystal structure of Cu deposition layer above the crystalline substrates Cu and Ni. The lower red line represents the Cu atoms depositing on Zr substrate. It shows amorphous structure, which is no long range order of the structure. The RDF of CuDepZr is almost identical to CuFilm(black line). There is a split in the second peak of the RDF of CuDepZr and CuFilm, which is characteristic of the amorphous solids [28]. The peak positions are: p1=2.55, p2=4.25, p3=4.95. The position ratio p2/p1 is 1.67, and p3/p1=1.94. The values are close to $\sqrt{3}$ and $\sqrt{4}$, which is a general character for metallic glass [29-31].

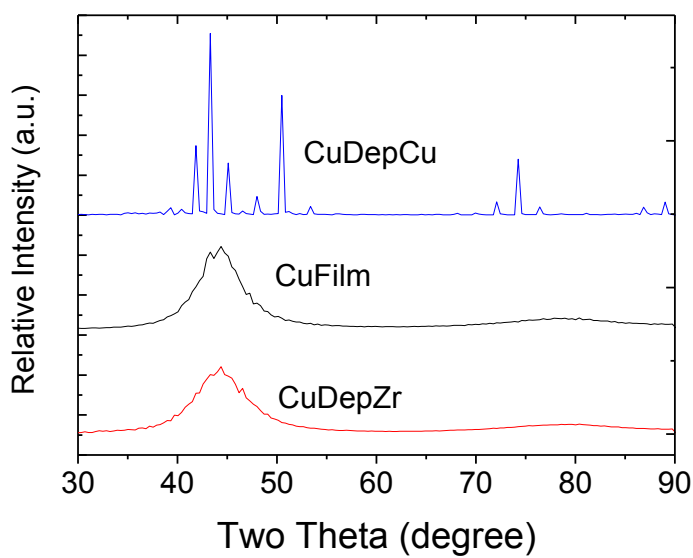

Figure 3. X-ray diffraction pattern of CuDepCu, CuFilm and CuDepZr. The Cu-K$_\alpha$ source with wavelength λ=1.541838 Å was applied in this work.

The X-ray diffraction (XRD) pattern [25] [26] [27] of CuDepCu, CuFilm and CuDepZr is in agreement of the RDF result. The crystal peaks for CuDepCu corresponds to the Cu fcc crystal structure. There is a broad diffraction maxima, which is characteristic for fully amorphous materials in the XRD pattern of CuDepZr and CuFilm. The amorphous peak is at position 44.38 degree. The XRD pattern of CuFilm is identical to CuDepZr..

From the result of RDF and XRD analysis, it is verified that the CuDepZr layer is completely amorphous. In contrast, the deposited Cu atoms on the crystal Cu (1 0 0) /Ni (1 0 0) substrate makes crystal structure immediately when the atom reaches the surface. Just like people join the queues automatically. It is because the crystal substrate generates regular potential landscape on the surface guiding the atoms to a position of the minimal energy. For the Zr crystal substrate, the first layer Cu-Zr potential disturbed the regular potential and the following atoms scatter disordered.

The deposition for the amorphous glass is at the temperature 100 K, when the temperature of 300 K is applied, the crystal structure is obtained, indicating besides suitable substrate choice, low temperature is also a key factor for the amorphous formation. The high temperature supply high kinetic energy to activate atoms override potential energy barrier from local minimum to crystallize.[32]

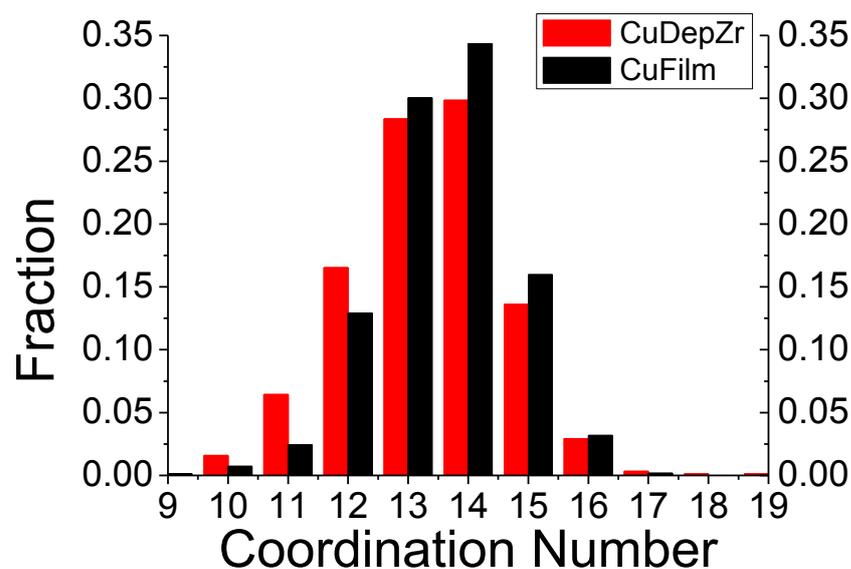

Figure 4 Coordination number distribution of CuDepZr (red) and Cufilm (black).

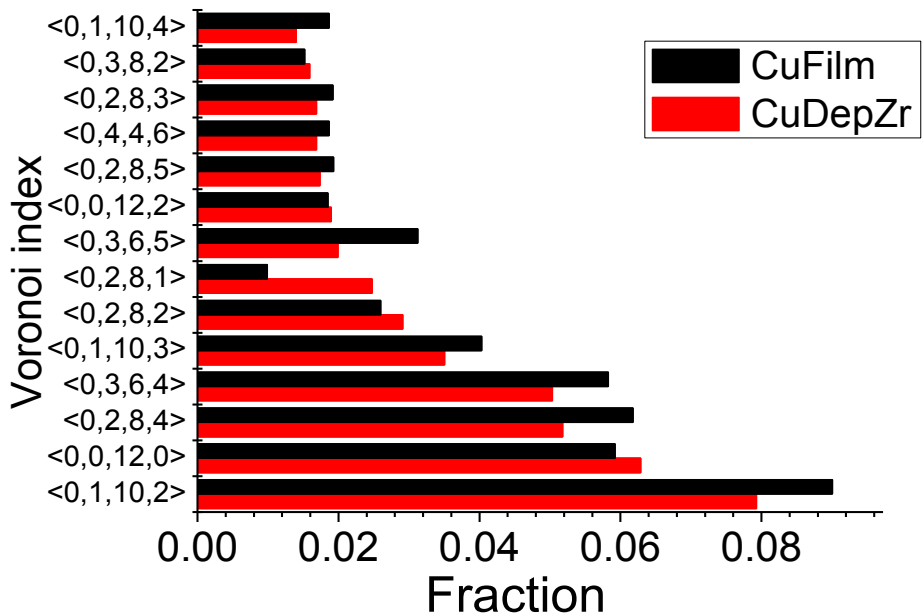

Figure 5 Distribution of Voronoi polyhedron of CuDepZr and CuFilm.

The RDF and XRD of CuDepZr and CuFilm are almost identical, but the microscopic atomic level structure shows obvious difference from the coordination number and Voronoi tessellation [24] analysis. The coordination number of a central atom is the number of its near neighbors. Figure 4 exhibits the distribution of coordination number of CuDepZr and CuFilm. For the CuFilm and CuDepCu, The population for coordination number of 10, 11 and 12 is higher for the CuDepZr than the CuFilm. Meanwhile the polyhedron with coordination number in excess of 13 possesses more population in the CuFilm. This means the CuDepZr prefers the low coordination numbers.

More detailed information of the atomic configuration can be obtained from the analysis of Voronoi tessellation[24]. The Schläfli notation[33] is applied in this work. The Voronoi index $n_i$ is the number of faces with i edges in a polyhedron (Wigner-Seitz cell). The population of polyhedrons with <$n_3$ $n_4$ $n_5$ $n_6$> is counted. The result is exhibited in figure 5. The Voronoi index <0 0 12 0> indicates full icosahedron. The population of icosahedron in CuDepZr is 6.3% and in CuFilm is 5.9%. This means the amorphous structure for both samples. The CuDepCu possesses obviously more polyhedrons with Voronoi index of <0 2 8 1>, <0 2 8 1> and <0 0 12 0> than the CuFilm. While the CuFilm contains more polyhedrons with Voronoi index of <0 1 10 2>,<0 2 8 4>, <0 3 6 4>, <0 1 10 3> and <0 3 6 5>. The other Voronoi index shows almost identical fraction in both samples.

**Conclusion:**

This simulation work proved the monatomic metallic glass could be obtained from deposition on crystalline substrate, only if the suitable deposition material and the substrate pair are selected. Except the suitable match, low temperature is also a key factor to prevent crystallization of the deposited atoms. Compare to the amorphous substrate, the crystal substrate is easy to get and it has smooth surface. The molecular dynamic simulation is a method for microstructural design, which would advance our understanding of liquids and glasses in general. This work provides a phenomenon for a better understanding of the metallic glass.

*Acknowledgements*

*Dr. Yang Yu acknowledges the financial support from the National Natural Science Foundation of China (Grant No.: 11247220) and the Foundation of Pre-Research of the Nanjing University of Information Science & Technology (Grant No:S8112103001). This work is also supported by the Natural Science Foundation of Jiangsu Province (No. BK20131428 )*

**Reference:**

1. Klement, W., R.H. Willens, and P.O.L. Duwez, *Non-crystalline Structure in Solidified Gold-Silicon Alloys.* Nature, 1960. **187**(4740): p. 869-870.
2. Kui, H.W., A.L. Greer, and D. Turnbull, *Formation of bulk metallic glass by fluxing.* Applied Physics Letters, 1984. **45**(6): p. 615-616.
3. Inoue, A., et al., *Preparation of 16 mm Diameter Rod of Amorphous $Zr_{65}Al_{7.5}Ni_{10}Cu_{17.5}$ Alloy.* Materials Transactions, JIM, 1993. **34**(12): p. 1234-1237.
4. Peker, A. and W.L. Johnson, *A highly processable metallic glass: Zr41.2Ti13.8Cu12.5Ni10.0Be22.5.* Applied Physics Letters, 1993. **63**(17): p. 2342-2344.
5. Schroers, J., *Bulk metallic glasses.* Physics Today, 2013. **66**(2): p. 32.
6. Greer, A.L., *Metallic glasses…on the threshold.* Materials Today, 2009. **12**(1–2): p. 14-22.
7. Liu, R.S. and S. Wang, *Anomalies in the structure factor for some rapidly quenched metals.* Physical Review B, 1992. **46**(18): p. 12001-12003.
8. An, Q., et al., *Synthesis of single-component metallic glasses by thermal spray of nanodroplets on amorphous substrates.* Applied Physics Letters, 2012. **100**(4): p. 041909.
9. Chen, F.F., et al., *Molecular dynamics study of atomic transport properties in rapidly cooling liquid copper.* The Journal of Chemical Physics, 2004. **120**(4): p. 1826-1831.
10. Rangsu, L., et al., *Formation and evolution properties of clusters in a large liquid metal system during rapid cooling processes.* Materials Science and Engineering: B, 2002. **94**(2–3): p. 141-148.
11. Zhong, L., et al., *Formation of monatomic metallic glasses through ultrafast liquid quenching.* Nature, 2014. **512**(7513): p. 177-180.
12. Schroers, J., *Condensed-matter physics: Glasses made from pure metals.* Nature, 2014. **512**(7513): p. 142-143.
13. Plimpton, S., *Fast Parallel Algorithms for Short-Range Molecular Dynamics.* Journal of computational physics, 1995. **117**(1): p. 1-19.
14. *LAMMPS Molecular Dynamics Simulator*. Available from: http://lammps.sandia.gov.
15. DÜNWEG, B. and W. PAUL, *BROWNIAN DYNAMICS SIMULATIONS WITHOUT GAUSSIAN RANDOM NUMBERS.* International Journal of Modern Physics C, 1991. **02**(03): p. 817-827.
16. Schneider, T. and E. Stoll, *Molecular-dynamics study of a three-dimensional one-component model for distortive phase transitions.* Physical Review B, 1978. **17**(3): p. 1302-1322.
17. Grønbech-Jensen, N. and O. Farago, *A simple and effective Verlet-type algorithm for simulating Langevin dynamics.* Molecular Physics, 2013. **111**(8): p. 983-991.
18. Grønbech-Jensen, N., N.R. Hayre, and O. Farago, *Application of the G-JF discrete-time*

*thermostat for fast and accurate molecular simulations.* Computer physics communications, 2014. **185**(2): p. 524-527.

19. Mendelev, M.I., D.J. Sordelet, and M.J. Kramer, *Using atomistic computer simulations to analyze x-ray diffraction data from metallic glasses.* Journal of Applied Physics, 2007. **102**(4): p. 043501.
20. Mendelev, M.I., et al., *Development of suitable interatomic potentials for simulation of liquid and amorphous Cu–Zr alloys.* Philosophical Magazine, 2009. **89**(11): p. 967-987.
21. Berk, O. and D. Sondan, *An optimized interatomic potential for Cu–Ni alloys with the embedded-atom method.* Journal of Physics: Condensed Matter, 2014. **26**(3): p. 035404.
22. OVITO. Available from: http://ovito.org/.
23. Alexander, S., *Visualization and analysis of atomistic simulation data with OVITO–the Open Visualization Tool.* Modelling and Simulation in Materials Science and Engineering, 2010. **18**(1): p. 015012.
24. Finney, J. *Random packings and the structure of simple liquids. I. The geometry of random close packing.* in *Proceedings of the Royal Society of London A: Mathematical, Physical and Engineering Sciences.* 1970: The Royal Society.
25. Coleman, S.P., D.E. Spearot, and L. Capolungo, *Virtual diffraction analysis of Ni [0 1 0] symmetric tilt grain boundaries.* Modelling and Simulation in Materials Science and Engineering, 2013. **21**(5): p. 055020.
26. Wilson, A.J.C. and V. Geist, *International Tables for Crystallography. Volume C: Mathematical, Physical and Chemical Tables. Kluwer Academic Publishers, Dordrecht/Boston/London 1992 (published for the International Union of Crystallography), 883 Seiten, ISBN 0-792-3-16-38X.* Crystal Research and Technology, 1993. **28**(1): p. 110-110.
27. Peng, L.-M., et al., *Robust Parameterization of Elastic and Absorptive Electron Atomic Scattering Factors.* Acta Crystallographica Section A, 1996. **52**(2): p. 257-276.
28. Zallen, R., *Frontmatter*, in *The Physics of Amorphous Solids*. 2007, Wiley-VCH Verlag GmbH. p. i-xi.
29. Bennett, C.H., *Serially Deposited Amorphous Aggregates of Hard Spheres.* Journal of Applied Physics, 1972. **43**(6): p. 2727-2734.
30. Liu, X.J., et al., *Metallic Liquids and Glasses: Atomic Order and Global Packing.* Physical review letters, 2010. **105**(15): p. 155501.
31. Wu, Z.W., et al., *Hidden topological order and its correlation with glass-forming ability in metallic glasses.* Nat Commun, 2015. **6**.
32. Heuer, A., *Exploring the potential energy landscape of glass-forming systems: from inherent structures via metabasins to macroscopic transport.* Journal of Physics: Condensed Matter, 2008. **20**(37): p. 373101.
33. Coxeter, H., *SM Regular Polytopes.* New York: Pitman Publishing Company, 1947.